
\def\sqr#1#2{{\vcenter{\vbox{\hrule height.#2pt
          \hbox{\vrule width.#2pt height#1pt \kern#1pt
           \vrule width.#2pt}
           \hrule height.#2pt}}}}

\hfill UMH-ULB-TH 05/93
\bigskip
\bigskip
\centerline {ENTROPY PRODUCTION FROM VACUUM DECAY}
\bigskip
\bigskip
\centerline{Ph. SPINDEL\footnote\dag{\rm Postal address: Universite de
Mons--Hainaut, Service de M\'ecanique et Gravitation, Facult\'e des Sciences,
15 av. Maistriau, B-7000 Mons, Belgium.}}
\centerline{Facult\'e des Sciences, Universit\'e de Mons--Hainaut}
\centerline{B-7000 Mons, Belgium.}
\bigskip
\bigskip
\centerline{R. BROUT\footnote\ddag{\rm Postal address: Universite libre de
Bruxelles, Service de physique Theorique, Campus Plaine C.P.225, Bd du
Triomphe,
B-1050 Brussels, Belgium}}
\centerline{Facult\'e des Sciences, Universit\'e Libre de
Bruxelles}
\centerline{B-1050 Brussels, Belgium}
\bigskip
\bigskip

{\sl Abstract}\/ : Using as dynamical variable the square of the radius of the
Universe, we solve analytically the Einstein equations in the framework of
Robertson-Walker models where a cosmological constant describing
phenomenologically the vacuum energy decays into radiation. Emphasis is put on
the computation of the entropy creation. \vfill


\endpage

Inflation in one form or another is commonly regarded as the  preadiabatic era
of our Universe. Nevertheless there does not exist at the present time a
consensus on how inflation stops\ref{E. Kolb and M. Turner,{\sl The Early
Universe}, Frontiers in Physics series, {\bf 69}, Addison-Wesley Publishing
Company (1990).}. In the present work we adopt a phenomenological approach in
which the cosmological constant $\Lambda$ is switched off. Our main interest
has
centered on estimating how the residual entropy depends on input parameters as
the Universe enters the adiabatic era. We recall that the presence of a
cosmological constant in the early (inflationary) Universe may be
interpreted as potential energy that is stocked inside the Universe (more
specifically, constant potential energy density). When inflation stops this
is then converted to energy carried by quanta (rest mass plus kinetic
energy). For definiteness and simplicity we have taken matter to be pure
radiation and we have neglected the effects of the trace anomaly. Our
analysis allows the possibility of a two stage switch-off wherein the initial
$\Lambda$ decays to a final value $\Lambda_\infty$. This latter subsequently
decays to zero, which process is then described by applying the analysis to
the case $\Lambda_\infty = 0$.

The motivation for our phenomenological analysis is not without ambiguity.
It has
become custumary in cosmology to refer to a phase transition scenario or to
something of the like using a (contrived) effective potential\ref{A. Linde,
{\sl Particle Physics and Inflationary Cosmology}, Harwood Academic (1990)}.
On the other hand there has been an important effort in recent years to show
that inflation stops for more fundamental reasons: infra--red fluctuations of
the conformal mode of gravity render de Sitter space unstable\ref{I.
Antoniadis, E. Mottola, J. Math. Phys.,{\bf 32}, 1037 (1991)}. An effort to
investigate the outcome of this instability has been furnished in further
work based on renormalization group techniques \ref {I. Antoniadis, E.
Mottola, Phys. Rev. D,{\bf 45}, 2013 (1992)} (limited unfortunately to the
conformal factor only) as well as numerical simulations\ref {I. Antoniadis,
P. Mazur and E. Mottola, preprint CPTH--A 214--1292 (1992)}. Though still
very tentative, this body of work gives preliminary indications that the
system tends to a fixed point which geometrically is flat space i.e. the
cosmological constant decays due to infra--red fluctuations of the
gravity--matter system. We may also quote the very early works\ref {R. Brout,
F. Englert and E. Gunzig, Ann. of Phys. {\bf 115},78 (1978); Gen. Rel. Grav.
{\bf 10}, 1 (1979)\hfill \break R. Brout, F. Englert
 and Ph. Spindel, Phys. Rev. Lett {\bf 43}, 417 (1979)\hfill\break
A. Casher, F. Englert, Phys. Lett. {\bf 104 B}, 117 (1981)} wherein inflation
was suggested as the condition of the preadiabatic universe brought about by
black hole production. Here it is the evaporation of black holes which would be
the responsible agent for the decay of the (effective) cosmological constant.
At
the present time our predilection is towards the body of work contained in
refs. [3--5     ]. For a qualitative review see Mottola
 \ref{ E. Mottola,{\sl
 Fluctuation-Dissipation Theorem in General Relativity and
 the Cosmological Constant}, Contributed talk at the Workshop on
``Physical Origins of Time Asymmetry'', Mazag\'on (Huelva), Spain, October 1
 1991.}.

In the absence of any reliable theory we have considered a simple model of
decay of the cosmological constant in radiation which has the virtue of
leading to equations which are solvable in simple terms.

We now present the  analytical solution of the model described above. We assume
that the geometry of the universe is  Robertson-Walker in character:
$$ds^2=-dt^2+a(t)\ d\Sigma_{k}^2$$
where as usual the values of $k$ correspond to closed ($k=+1$),
open ($k=-1$) or the limiting flat  ($k=0$) spatial sections. The energy
momentum scheme is given by the
sum of  two pieces, one describing the decaying cosmological constant $$ \/
_{v}T_{\mu}^{\nu} = \Lambda (t) \delta _{\mu}^{\nu}$$ with $$\Lambda (t)=
\Lambda _0 e^{-{t\over \tau}} + \Lambda _{\infty}$$ corresponding to a vacuum
energy decay, starting at $t=0$ with a rate $\tau^{-1} $ and evolving from
$\Lambda _0+\Lambda _{\infty}$ to a residual value  $ \Lambda _{\infty}$
\REF\prigo{E. Gunzig,  J. G\'eh\'eniau, I. Prigogine,Nature
{\bf 330}, 621 (1987)\hfill\break E. Gunzig, Nature {\bf 337}, 216 (1988)}
\footnote {*}{ The case  $\tau=0$, $ \Lambda _{\infty}=0$, $k=0$ has been
discussed previously\refend .}.
The
other, more standard, describes highly relativistic matter in thermal
equilibrium: $$ \/ _{r}T_{\mu}^{\nu} = \bigl(\rho(t) + p(t)\bigr) \delta
_{\mu}^0\delta ^{\nu}_0- p(t)\delta_{\mu}^{\nu}$$
with $\rho = 3p$.

 The nice
trick to solve Einstein equations is to decouple the radiation part by  using
as
variable the square of the radius function: $z=a^2$.This leads to the equations
(written in Planck units such that $8\pi G =1$)
 $$ {1\over 4} ({\dot z \over
z})^2\ +\ {k\over z}\ =\ {1\over 3}\bigl(\Lambda(t)\ +\ \rho (t)\bigr)\qquad
,\eqn \goo$$
$$ \ddot z \/ -\/ {4\over 3}\Lambda(t) \/ z\ = \/ -2k \qquad .\eqn
\dyn$$ The Bianchi identities reduce to the entropy balance equation~: $$
(\rho\/z^2)\dot {\strut}\ = \ {\Lambda_{0}\over \tau}z^2\/ e^{-{t\over
\tau}} \ =
-\dot \Lambda(t) \/ z^2\qquad .\eqn \balance $$

Introducing dimensionless variables~: $$ \xi\ = \ -{t \over 2\tau}\quad
,\quad\zeta\ =\ {z\over 4 \tau^2}\quad ,$$ $$ \lambda_{0}\ =\ {16\over
3}\tau^2\/ \Lambda_{0}\quad , \quad \lambda_{\infty}\ =\ {16\over 3}\tau^2\/
\Lambda_{\infty}\quad , \quad \sigma_{0}\ =\ {16\over 3}\tau^2\/ \rho (0)$$
allows us to write eq. \dyn \ as $$ {{d^2\zeta}\over{d\xi ^2}} -(\lambda _{0}
e^{2\xi}+\lambda _{\infty})
 \zeta\ = -2k \qquad . \eqn \dynd$$ The general solution of this equation is
expressible  in terms of modified Bessel functions as $$\zeta\ = \ K_\nu(\varpi
e^\xi)\/\bigl(\alpha+2k\int^\xi_0 I_\nu(\varpi e^x)\/dx\bigr) \/+\/
I_\nu(\varpi
e^\xi)\/\bigl(\beta-2k\int^\xi_0 K_\nu(\varpi e^x)\/dx\bigr) \eqn \gensol$$
where the
constants $\alpha $ and $\beta$ are determined by the initial conditions~:
$$\eqalign {\alpha\  &=\ \varpi I^{\prime}_{\nu}(\varpi)\/\zeta _{0}\/-\/
I_{\nu}(\varpi)\/(\zeta^2_0 (\lambda_0+\lambda_\infty +\sigma_0)-4k
\zeta_0)^{1\over 2}\quad ,\cr \beta\ &=-\varpi K^{\prime}_{\nu}(\varpi)\/\zeta
_0 + K_{\nu}(\varpi)\/(\zeta^2_0 (\lambda_0+\lambda_\infty +\sigma_0)-4k
\zeta_0)^{1\over 2}\cr}$$ and we have set $$\nu\ =\ \sqrt
\lambda_{\infty}\quad {\rm and}\quad \varpi \ =\ \sqrt \lambda_{0}\qquad .$$
 A prime
denotes the derivative of the function with respect to its  arguement.

Using the expansions of the Bessel functions near the origin, valid at
asymptotically large times, or directly from eq. \balance , we see that $z$
grows asymptotically as $\exp (\nu t)$. From eq. \dyn \  we conclude then
that the production of entropy per comoving volume $S\propto\/ (\rho
z)^{3/4}$ diverges if $\nu \ge 1$ i.e. $\tau^2\Lambda_\infty \ge {3\over
16}$. The  asymtotic value
 of the entropy produced $S_\infty$ is calculated directly from the first
Einstein
equation (eq. \goo ):
 $$ \eqalign{S_{\infty}^{4\over 3} &\propto 4\tau^2[({d\zeta \over
d\xi})^2-\varpi ^2e^{2\xi} \zeta ^2 - \nu ^2 \zeta ^2 + 4 k \zeta]\cr
\/&={4\over 3}\rho(t)\ z^{2}\qquad .}$$
When $0 < \nu <1 $ we obtain for large values of $t$ : $$S^{4\over 3}\propto
4\tau^2\bigl(({\pi \nu \over \sin \nu \pi} n^2-2\nu n\/ q
 +{4 k^2\over \nu ^2})+(n\/\Gamma (\nu)\/ 2^{(\nu-1)})^2\/e^{{(\nu-1)\over
\tau}t})\bigr) $$
 where
$$n=\alpha -2k\int^0_{-\infty}I_{\nu}(\varpi e^\xi)d\xi$$
 and
$$q=\beta+2k\int^0_{-\infty}(K_{\nu}(\varpi e^\xi)-{{\Gamma(\nu)}\over
{2}} ({{\varpi}\over {2}})^{-\nu}e^{-\nu \xi})d\xi\ -k{{\Gamma(\nu)}\over
{\nu}}
({{\varpi}\over {2}})^{-\nu}$$
while for $\nu=0$ we have in the limit where $t$ goes to $\infty$:
 $$S_{\infty}^{4\over 3}\propto 4\tau^2[{\tilde n}^2+4k({\tilde
q}-{\tilde n}(\log {\varpi \over 2}+\gamma)]$$
where
$$\eqalign{ {\tilde
n}&=\alpha -2k\int^0_{-\infty}(I_{0}(\varpi e^\xi)-1)d\xi\qquad ,\cr
{\tilde q}&=\beta +2k\int^0_{-\infty}(K_{0}(\varpi e^\xi)+\xi+
\log {\varpi \over
2}+\gamma)d\xi\qquad ,}$$
and $ \gamma=.577\dots$ is the Euler-Mascheroni's
constant.

{}From these results the following conclusions may be inferred:
\item{1.} If
$\Lambda _{\infty} \neq 0$ the amount of total entropy produced per unit
 covolume
depends on a balance between the rate of decay of the vacuum energy and the
rate of expansion of the Universe.
\item{2.} If $\Lambda _{\infty} = 0$ the
amount of total entropy produced per unit covolume still depends on the rate of
decay
but only very slow decay will leads to large entropy. For instance if we
restrict
ourselves to the special case $k=0$ and $\Lambda_{\infty}=0$, we
 obtain for the entropy produced (neglecting all variations in the number of
relativistic degrees of freedom):
$$S_{\infty}\propto\Bigl(\sqrt{\rho_0+\Lambda_0}\ I_{0}(\varpi)
-\sqrt{\Lambda_0}\ I_{1}(\varpi)\Bigr)^{3/2}$$
 whose limiting forms up to a multiplicative factor are
$$S_{\infty}\approx \Bigl(\sqrt{\rho_0+\Lambda_0}
-\sqrt{\Lambda_0}\Bigr)^{3/2}\eqn \asym$$
 for fast decay: $\tau^2\Lambda_0<<1$, and
$$S_{\infty}\approx {e^{2\sqrt{3\Lambda_{0}}\tau}\over (8\pi\tau
\sqrt{\Lambda_{0}/3})^{3/2}}
\biggl[\Bigl(\sqrt{\rho_0+\Lambda_0}
-\sqrt{\Lambda_0}\Bigr)+{\sqrt{3}\over
32\tau}\Bigl(\sqrt{1+{\rho_0\over\Lambda_0}}
+3\Bigr)\biggr]^{3/2}\eqn \asymf$$
for slow decay: $\tau^2\Lambda_0>>1$ .\hfil \break
\FIG\un{Logarithmic plots of the asymptotics values of the entropy
created as function of the decay rate under the conditions
$\Lambda_{\infty}=0$ and $k=0$.}Fig.\un\ shows
various logarithmic plots of ${3\over2}\log(\sqrt{\rho_0+ \Lambda_0}\
I_{0}(\varpi) -\sqrt{\Lambda_0}\ I_{1}(\varpi)\ \propto\ \log S_\infty\ .$
Both its decreasing
 behaviour for small values of $\tau$ (see eq. \asym) and its exponential
blow up
 for large value (see eq.\asymf)
 appear clearly and reflect the competition between the r\^ole of a rapid
decrease of
 $\Lambda$ contributing directly to increase the entropy and a slow decrease
 which contribute also, but indirecty via the incease of the covolume in
which entropy is generated (see eq.\balance). It is also seen that large values
of $\tau $ result in an exponentially large entropy production, so
that this simple mechanism of conversion of vacuum energy to radiation indeed
is
efficint as a source of cosmological entropy.
\item{3.} The reader may be perplexed in that in the limit $\tau\ \rightarrow\
\infty$ one has $\dot \Lambda =0$ and hence no entropy production at all. But
we have just seen that large values of $\tau$ give rise to exponentially large
entropy production. The point is here that the entropy so produced takes a
time of
$\cal{O}(\tau)$ (See the plateau of \FIG\dx {Logarithmic plots of the
 entropy produced as function of time under the
conditions $\Lambda_{\infty}=0$ and $k=0$. Note the first plateau
representing the time needed for the entropy to be significantly
produced.}Fig. \dx . It is easy to show from eq. \gensol \ that this plateau
is a general feature of entropy production). In consequence we
 finds the two limits $\tau\ \rightarrow\ \infty$ and $t\ \rightarrow\
\infty$ do not commute.
\item{4.}The plateau of Fig. \dx \ is due to the balance of cooling the
radiation due to expansion and heating due to its acquisition of energy from
the decay of  $\Lambda $. The subsequent rise is called reheating and is akin
to the discussion of reheating given in refs [1,2] based on the phase
transistion or effective potential scenarios. This region is radiation
dominated (i.e. radiative energy $>$ energy carried by the cosmological
constant).
 \item{5.}Decay of $\Lambda $ which is other than exponential can be
studied using the same technique.In this way one can show that when $\Lambda $
decreases faster than $t^{-2}$, say as $t^{-\kappa}$,  $z$ grows as $t^{2}$ if
$k=-1$ (as $t$ if $k=0$) and the entropy produced will still diverges if
$\kappa
> 3$ (if $\kappa > 1$ in the unphysical case where $k=0$). If $\Lambda $
>decreases
more slowly than $t^{-2}$ then the radius of the Universe blows up
exponentially~:
$$z \propto t^{\kappa \over 4} e^{2 \sqrt{\Lambda_{0}\over 3}
{t^{(1-\kappa /2)}
\over{(1-\kappa/2)}}}$$
 \item{6.} The trick which is used both to decouple the radiation
and to linearize the second order trace equation is of more general
applicability
than we have displayed here. Given a particular Robertson-Walker cosmological
solution characterized by a non radiative energy source $\rho _{m}$, one may
generate (by quadratures) an other solution which contains radiation in
addition
to $\rho _{m}$.
\ack We would like to thank Jean-Marie Fr\`ere and Renaud Parentani for
numerous stimulating discussions.
 \refout
\figout
\end